\documentclass[aps,prl,showpacs,twocolumn]{revtex4}
\usepackage{graphicx}
\usepackage{amsmath}

\def\etal{\emph{et al.}}
\def\insitu{\emph{in situ }}
\def\vs{\emph{vs. }}
\def\ie{\emph{i.e.}}

\begin{document}

\title{Quantitative measurements of the thermal resistance of Andreev interferometers}
\author{Z. Jiang and V. Chandrasekhar}
\affiliation{Department of Physics and Astronomy, Northwestern University, Evanston, IL 60208, USA}
\date{\today}

\begin{abstract}
Using a local thermometry technique, we have been able to quantitatively measure the thermal resistance $R^T$ of diffusive Andreev interferometers. We find that $R^T$ is strongly enhanced from its normal state value at low temperatures, and behaves non-linearly as a function of the thermal current through the sample. We also find that the $R^T$ oscillates as a function of magnetic flux with a fundamental period corresponding to one flux quantum $\Phi_0=h/2e$, demonstrating the phase coherent nature of thermal transport in these devices. The magnitude of $R^T$ is larger than predicted by recent numerical simulations.
\end{abstract}

\maketitle

A diffusive normal metal (N) in proximity to a superconductor (S) in a mesoscopic hybrid device also acquires superconducting properties by the process of Andreev reflection \cite{andreev}: at temperatures well below the gap of the superconductor, $k_BT<<\Delta$, an electron in the normal metal cannot be transmitted through the NS interface, but is reflected as a coherent hole with the simultaneous generation of a Cooper pair in the superconductor. The electrical transport properties of such proximity-coupled systems have been extensively investigated both experimentally and theoretically in the last decade \cite{ex1,ex2,th1,th2,th3}. More recently, the thermal transport properties have attracted much theoretical interest \cite{bezuglyi,jiang1,heikkila,chandrasekhar}, following measurements of the phase-dependent thermopower of Andreev interferometers \cite{eom,dikin1,parsons}. In elastic-scattering dominated normal-metal systems, the ratio of the electrical to the thermal resistance is proportional to the temperature, the so-called Wiedemann-Franz (WF) law \cite{wiedemann}. However, theoretical studies indicate that the thermal resistance of a normal metal in the proximity regime is strongly enhanced \cite{bezuglyi,jiang1,chandrasekhar}. The WF law, which is widely valid in a disordered metal system, is no longer correct for proximity-coupled systems. This topic was first explored experimentally by Dikin {\etal} \cite{dikin2}. However, the sample in that experiment had two superconducting elements directly in the path of the thermal current, so that the reduced thermal conductance may have been due to the well-known suppression of thermal conductance in a conventional superconductor \cite{dikin2,bezuglyi} rather than a proximity effect phenomenon. 

In this Letter, we report measurements of the temperature and magnetic field dependence of the thermal resistance $R^T$ of Andreev interferometers (consisting of a hybrid loop with one superconducting arm and one normal-metal arm) without superconductors in the thermal current path, and therefore in the true proximity regime.  As predicted by theory \cite{bezuglyi,jiang1,chandrasekhar}, we find that $R^T$ of a proximity-coupled normal metal is enhanced by as much as one order of magnitude from its normal state value at low temperatures. $R^T$ also oscillates as a function of the external magnetic flux, with a fundamental period of one superconducting flux quantum $\Phi_0=h/2e$.  These oscillations demonstrate the phase coherent nature of the thermal current in this system. Furthermore, $R^T$ is a strongly nonlinear function of the thermal current $I^T$ even at thermal currents as small as a few femtowatts.

\begin{figure}[b]
\includegraphics[width=8.6cm]{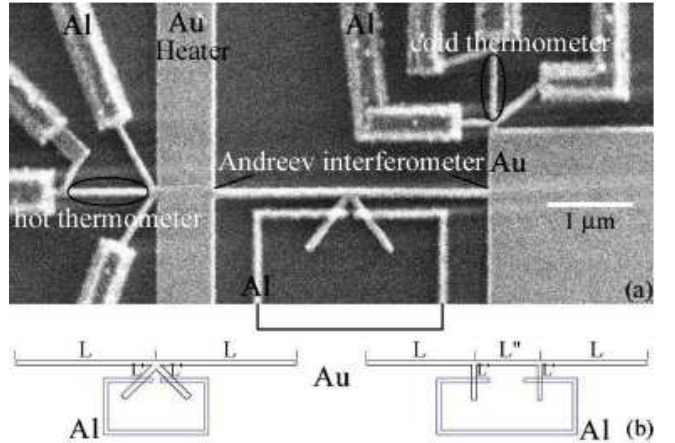}
\caption{(a) Scanning electron microscope (SEM) image of one of the devices. The device consists of three parts: (1) On the left, a metallic heater line with a thermometer. (2) On the right, a large normal-metal pad with another thermometer. (3) In the middle, a `house' Andreev interferometer. (b) Schematic of Andreev interferometers we measured with two different geometries: (left) `house' and (right) `parallelogram.'}
\label{fig1}
\end{figure}

Experimentally, the thermal resistance is defined as the ratio of the temperature differential $\Delta T$ across the sample to the thermal current $I^T$ through the sample, \ie, $R^T\equiv \Delta T/I^T$, under the condition that no electrical current flows through the sample ($I=0$). Hence, in order to obtain $R^T$, our devices should be designed
so that $I^T$ and  $\Delta T$ can be measured directly.  Figure 1(a) shows a scanning electron microscopy (SEM) image of one of the devices we measured.  There are two layers of metal on this device. The first layer is a 50 nm thick Au film, followed by a 100 nm thick Al film deposited on top of the Au in a second level of lithography, after an \insitu oxygen
plasma etch was used to clean the Au surfaces in order to obtain good NS interfaces.  The device consists of three parts:  (1) In the center is the sample itself, which in this case is an Andreev interferometer in the `house' configuration (in the terminology introduced in Ref. \cite{eom}). The Andreev interferometer includes a one-dimensional (1D) normal-metal (Au) wire and a superconducting (Al) loop.  (2) On the left is a heater, which is essentially a 0.68-$\mu$m-wide Au line connected electrically to the sample.  By passing  a direct (dc) current through this heater, one can raise the electron temperature at one end of the sample.  (3) On the right is a large,  normal-metal pad that serves as the cold end of the sample.  Attached to both the heater and the cold normal-metal pad are thermometers that measure the local electronic temperature.  The normal-metal heater is approximately 25 $\mu$m long, much longer than the
inelastic electron scattering length, so that one can define a effective local electronic temperature in the middle of the heater that can be measured by the thermometer on the left, which we denote the `hot' thermometer (the thermometer attached to the other end of the sample is the `cold' thermometer).  The design and operation of these thermometers
have been described in detail elsewhere \cite{jiang2}, and will not be discussed here.    Connections to both end of the heater are made through superconducting Al contacts to reduce the heat flow.  Two additional probes at each end allow us to measure the total four-terminal differential resistance of the heater for any value of the dc current through it, thereby determining the power generated in the heater.   In total, five devices were measured, in two different configurations, the `house'  and `parallelogram' geometries shown in Fig. 1(b).  Data for three of these devices are discussed in this paper.

\begin{figure}[b]
\includegraphics[width=8cm]{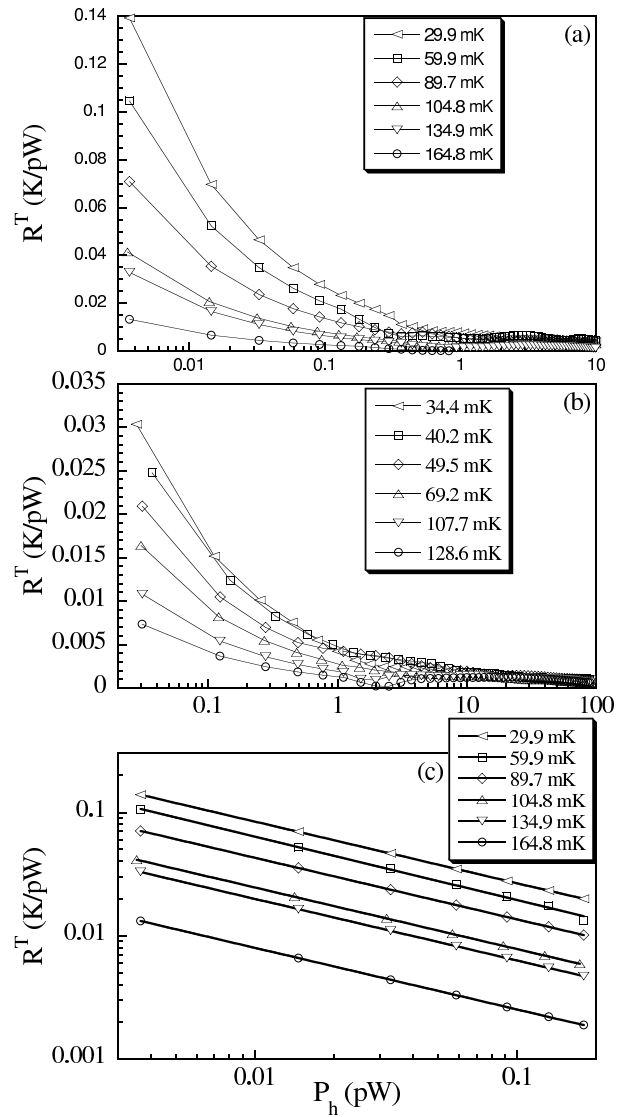}
\caption{Thermal resistance $R^T$ as a function of the heater power $P_h$ at six different base temperatures for (a) the `house' and (b) the `parallelogram' interferometers. (c) $R^T$ \vs $P_h$ at small values of $P_h$ for the `house' geometry.  The solid lines are fits to the functional form $R^T\propto\sqrt{1/P_h}$.  For the `house' interferometer, the distances from either side of the sample and from the NS interfaces to the center node are (referring to Fig. 1(b)) $L=1.55$ $\mu$m and $L'=0.29$ $\mu$m respectively.    For the `parallelogram,' $L=1.19$ $\mu$m, $L'=0.24$ $\mu$m, and $L"=0.76$ $\mu$m.}
\label{fig2}
\end{figure}

To determine the thermal resistance, the electrical resistances of both the hot and the cold thermometer are measured as a function of the temperature of the dilution refrigerator mixing chamber, with no dc current through the heater.  The resistances of both thermometers are then measured as a function of the dc current $I_h$ through the heater, with the base temperature of the dilution refrigerator fixed.  The two measurements are then cross-correlated to obtain the effective electron temperature as measured by the hot and cold thermometers as a function of $I_h$; a simultaneous measurement of the differential resistance of the heater determines the equivalent power $P_h$ generated in the heater.
The difference between the temperatures measured by the two thermometers gives $\Delta T$.  Since all connections to the heater except the connection to the Andreev interferometer are made via superconducting contacts whose thermal conductance is negligible at temperatures far below the transition temperature, the power generated in the heater can
only flow out through the Andreev interferometer, or through the substrate.  At temperatures below about 200 mK, the electron-phonon coupling in the normal metal is very weak, so that thermal leakage to the substrate can be ignored \cite{jiang2}.  In addition, while our earlier devices (such as the parallelogram interferometer discussed in this
paper) were fabricated on oxidized Si substrates, more recent devices (such as the house interferometer shown in Fig. 1(a)) are fabricated on 50 nm thick Si$_3$N$_4$ substrates.  In these samples, the thermal leakage to the substrate is even smaller, so that in principle, the valid range of measurement can be extended to higher temperatures.  In practice, the temperature range is restricted by the sensitivity of the thermometers.  Since all the power generated in the heater flows through the Andreev interferometer, the thermal current through the sample $I^T$ is simply given by the measured value of $P_h$,   and the thermal resistance $R^T$ is given by $\Delta T/P_h$.  This measurement is repeated at different values of the base temperature of the refrigerator to obtain $R^T$ as a function of temperature.
  
Figure 2(a) and (b) show the thermal resistance $R^T$ of a `house' and a `parallelogram' interferometers at six different temperatures as a function of the power of the heater $P_h$.   As shown in Fig. 2, the qualitative behavior of $R^T$ as a function of $P_h$ is similar for the two geometries. In order to experimentally define a thermal resistance in the linear response regime, $R^T$ should approach a limiting value as $I^T\rightarrow0$.  However, $R^T$ continues to change as a function of $P_h$ (or equivalently, $I^T$) even at a  heater power of a few femtowatts. At low values of $P_h$, $R^T\propto \sqrt{1/P_h}$, as shown in Fig. 2(c). This power law dependence of $R^T$ on $P_h$ is valid for all samples and at all temperatures measured.  Theoretically, of course, a linear response $R^T$ can be defined; however, numerical simulations for these devices based on the quasiclassical theory \cite{jiang1} also find that $R^T\propto \sqrt{1/P_h}$ for intermediate values of $P_h$.  These simulations show that the linear response regime is approached for $P_h\lesssim $10 fW, comparable to the minimum values used in the experiments.

\begin{figure}[b]
\includegraphics[width=8.6cm]{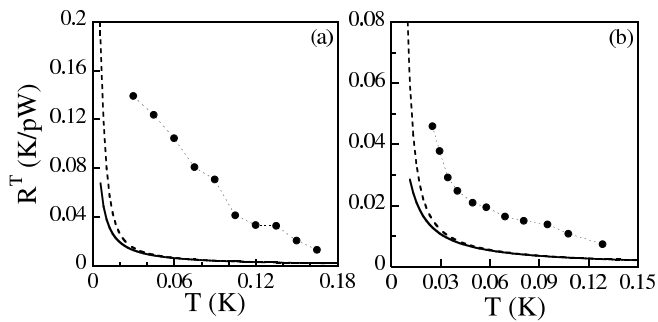}
\caption{The solid symbols are the thermal resistance $R^T$ of (a) the `house' interferometer and (b) the `parallelogram' interferometer as a function of the mixing chamber temperature, determined from the data of Fig. 2 at the lowest heater power applied in the experiments. The dotted lines are guides to the eye.  The solid lines represent the thermal resistance of equivalent normal-metal wires, estimated using the Wiedemann-Franz law, and the measured normal-state electrical resistance of the interferometers.  The dashed lines represent theoretical calculations of the thermal resistance of the Andreev interferometers, using the experimental parameters for the samples, as described in the text.}
\label{fig3}
\end{figure}

Given the nonlinear behavior of $R^T$ on $P_h$,  we shall experimentally define the linear response thermal resistance as the value of $R^T$ at the lowest heater power measured. This corresponds to $P_h=3.7$ fW for the `house' interferometer and $P_h=31$ fW for the `parallelogram' interferometer of Fig. 2 respectively.  The solid symbols in Figs. 3(a) and 3(b) show the resulting thermal resistance as a function of temperature for the `house' and `parallelogram' interferometers.  For comparison, we also show the expected thermal resistance for an equivalent normal-metal sample, calculated using the Wiedemann-Franz law from the measured normal-state resistance of the wire, using the textbook value for the Lorenz number.  For both samples, $R^T$ increases rapidly with decreasing temperature.  For the `house' thermometer,  $R^T$ is larger than the thermal resistance for an equivalent normal system by almost an order of magnitude at the lowest temperature.  For the `parallelogram' interferometer, at first sight, the increase is not as large; however, it must be noted that $R^T$ for this sample was inferred at a heater power of 31 fW.  If we extrapolate $R^T$ to a value of 3.7 fW as for the `house' interferometer, we obtain values of $R^T$ of the same order of magnitude as in Fig. 3(a).

Depending on the dimensions of the sample, the electrical resistance of proximity coupled normal-metals can show `reentrant' or non-monotonic behavior, where the resistance first decreases below the transition of the superconductor, but reaches a minimum and then starts increasing as the temperature is decreased further \cite{ex2}.  The non-monotonic
behavior is associated with the competition between two effects, a decrease in the resistance due to pair correlations induced in the normal metal, and an increase in the resistance due to a decrease in the density of states $N(E)$ near the Fermi energy $E_F$.  At any finite temperature, the effect of pair correlations is greater, but at $T=0$ there is
expected to be an exact cancellation for systems without interactions, resulting in the system regaining its normal-state resistance at $T=0$. Theoretically, the thermal resistance of the `house' interferometer is expected to be influenced only by the decrease in $N(E)$, and hence increases monotonically as the temperature decreases.  Unlike a
superconductor, however, $N(E)$ does not go to zero as $T\rightarrow0$, but saturates at a value that depends on the dimensions of the sample and the transparency of the NS interfaces.  As $T\rightarrow0$, $N(E)$ is small but finite; the quasiparticles occupying the levels in the pseudogap contribute to the thermal conductance, leading to an enhanced
thermal resistance, but one that still varies inversely with $T$ according to the Wiedemann-Franz law.  Numerically,  $R^T$ approaches this limiting behavior below a temperature corresponding to approximately $0.1 E_c/k_B$, where $E_c=\hbar D/L_0^2$ is the correlation energy \cite{jiang1}.  Here $D$ is the electronic diffusion coefficient in the
normal metal, and $L_0$ is the length from one end of the interferometer to one of the NS interfaces.  In the `house' interferometer of Figs 2 and 3, for which the diffusion constant $D$ = 208 cm$^2$/sec, and $L_0=1.84$ $\mu$m, $0.1 E_c/k_B \sim$ 4.7 mK, below the temperature range of the experiment.  Equivalent parameters for the `parallelogram'
interferometer are  $D$ = 127 cm$^2$/sec and $L_0$ = 1.43 $\mu$m , giving a similar saturation temperature of $0.1 E_c/k_B \sim$ 4.7 mK.  The dashed lines in Fig. 3(a) and 3(b) show the result of the numerical simulations of the thermal resistance of the `house' and `parallelogram' interferometers respectively, with the parameters given above, and
assuming perfectly transparent NS interfaces \cite{jiang1}.   The theoretical predictions show significant deviations from the normal state thermal resistance only at temperatures below 20-30 mK, while the experimental $R^T$ is already larger than the normal state thermal resistance at temperatures about an order of magnitude higher. We believe this deviation might be due to an intrinsic mesoscopic effect not restricted to NS devices associated with the long length scales required to equilibrate the energy of the quasiparticles in mesoscopic systems.

\begin{figure}[t]
\includegraphics[width=8.6cm]{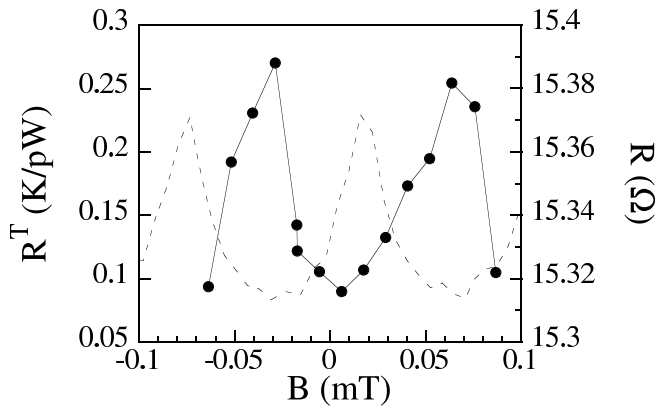}
\caption{The solid symbols are the thermal resistance of a `house' interferometer at different magnetic fields, measured at $T=40$ mK. The solid line is a guide to the eye.  The dashed line is the resistance of the same sample as a function of the magnetic field at $T=400$ mK.}
\label{fig4}
\end{figure}

The phase dependent nature of transport in mesoscopic NS structures means that the thermal resistance of an Andreev interferometer should oscillate periodically with an externally applied magnetic field, with a fundametal period corresponding to a flux $h/2e$ through the area of the interferometer loop \cite{bezuglyi,jiang1}. Detailed simulations in Ref. \cite{jiang1} show that the `house' interferometer has a larger oscillation amplitude of the thermal resistance compared with the `parallelogram' interferometer, hence we focused on the former in measuring oscillations of $R^T$.  Experimentally, measuring the oscillations in $R^T$ turns out to be a difficult proposition.  Ideally, we would like to bias the sample at a specific heater power and look at the variations in the temperature measured by the local electron
thermometers as we sweep the magnetic field.  However, at a heater power of 3 fW, the maximum variation in the temperature difference would be on the order of 0.5 mK (as can be seen from Fig. 2(a)), below our experimental resolution.  One could apply a larger amount of power to the heater; however, the maximum variation in $R^T$ drops dramatically at larger values of $P_h$.  Consequently, in order to measure the variation of $R^T$ with external magnetic field $B$, we use the same technique as for the temperature dependence:  at a fixed external magnetic field $B$,
we measure the thermal resistance as a function of heater power $P_h$, and take the value at the lowest measured heater power as the value of $R^T$ at that value of magnetic field $B$.  Figure 4 shows the resulting $R^T$ for a third Andreev interferometer device in the `house' configuration, measured at $T$=40 mK, at $P_h$= 2.1 fW.  For comparison, we also show oscillations of the resistance of the interferometer measured at $T$=400 mK.  It can be seen that $R^T$ oscillates
periodically as a function of  $B$, with a period corresponding to one superconducting flux quantum $\Phi_0=h/2e=0.092$ mT through the interferometer loop.  Both the electrical resistance $R$ and $R^T$ have the same symmetry with respect to $B$.  Since $R$ is known to be symmetric with respect to $B$, this means that $R^T$ is also symmetric with respect to $B$, as predicted by theory (the offset seen in the data is due to the remanent field in the external superconducting solenoid).

In summary, we have measured the thermal resistance of Andreev interferometers in two different geometries. We find that the measured thermal resistance of all the samples is enhanced at low temperatures and deviates from the values estimated from the WF law for equivalent normal-metal systems. In addition, the measured thermal resistance shows
strong nonlinear behavior with respect to the thermal current $I^T$. At small values of $I^T$, $R^T\propto \sqrt{1/I^T}$.   Finally, we have observed that the thermal conductance oscillates periodically as a function of the applied magnetic flux with a fundamental period corresponding to $\Phi_0=h/2e$. Our results are qualitatively consistent with recent numerical simulations \cite{jiang1}.

This work is supported by the NSF through grant number DMR-0201530.

\end{document}